\documentclass[aps,prb,twocolumn,showpacs,preprintnumbers,amsmath,amssymb,superscriptaddress]{revtex4}%

\usepackage{graphicx}%
\usepackage{dcolumn}
\usepackage{amsmath}
\usepackage{color}
\usepackage{bm}% bold math

\makeatletter
\def\btt#1{\texttt{\@backslashchar#1}}%
\DeclareRobustCommand\bblash{\btt{\@backslashchar}}%
\makeatother

\begin{document}

\preprint{PREPRINT (\today)}

\title{Correlation between oxygen isotope effects on the transition temperature and
the magnetic penetration depth in high-temperature superconductors
close to optimal doping }

\author{R.~Khasanov}
 \affiliation{Physik-Institut der Universit\"{a}t Z\"{u}rich,
Winterthurerstrasse 190, CH-8057, Z\"urich, Switzerland}
\affiliation{Laboratory for Muon Spin Spectroscopy, Paul Scherrer
Institut, CH-5232 Villigen PSI, Switzerland}
\author{A.~Shengelaya}
 \affiliation{Physics Institute of Tbilisi State University,
Chavchadze 3, GE-0128, Tbilisi, Georgia } %%
\author{K.~Conder}
\affiliation{ Laboratory for Developments and Methods, Paul Scherrer
Institut, CH-5232 Villigen PSI, Switzerland }
\author{E.~Morenzoni}
\affiliation{Laboratory for Muon Spin Spectroscopy, Paul Scherrer
Institut, CH-5232 Villigen PSI, Switzerland}
\author{I.M.~Savi\'c }
\affiliation{Faculty of Physics, University of Belgrade, 11001
Belgrade, Serbia and Montenegro}
\author{J.~Karpinski}
\affiliation{Solid State Physics Laboratory, ETH 8093 Z\"urich,
Switzerland}
\author{H.~Keller}
\affiliation{Physik-Institut der Universit\"{a}t Z\"{u}rich,
Winterthurerstrasse 190, CH-8057, Z\"urich, Switzerland}

\begin{abstract}
The oxygen-isotope ($^{16}$O/$^{18}$O) effect (OIE) on the in-plane
magnetic penetration depth $\lambda_{ab}(0)$ in optimally-doped
YBa$_2$Cu$_3$O$_{7-\delta}$  and La$_{1.85}$Sr$_{0.15}$CuO$_4$, and
in slightly underdoped YBa$_2$Cu$_4$O$_8$ and
Y$_{0.8}$Pr$_{0.2}$Ba$_2$Cu$_3$O$_{7-\delta}$ was studied by means
of muon-spin rotation. A substantial OIE on $\lambda_{ab}(0)$ with
an OIE exponent $\beta_{\rm O}=-{\rm d} \ln\lambda_{ab}(0)/{\rm d}
\ln M_{\rm O}\approx - 0.2$ ($M_{\rm O}$ is the mass of the oxygen
isotope), and a small OIE on the transition temperature $T_c$ with
an OIE exponent $\alpha_{\rm O}=-{\rm d} \ln T_{c}/d \ln M_{\rm
O}\simeq 0.02$ to 0.1 were observed. The observation of a
substantial isotope effect on $\lambda_{ab}(0)$, even in cuprates
where the OIE on $T_c$ is small, indicates that lattice effects play
an important role in cuprate HTS.
\end{abstract}
\pacs{76.75.+i, 74.72.-h, 82.20.Tr, 74.25.Kc}

\maketitle

\section{Introduction}
The observation of unusual isotope effects in cuprate
high-temperature superconductors (HTS) on the transition temperature
$T_c$ \cite{Batlogg87a,Franck91,Franck94,Zech94} and on the zero
temperature in-plane magnetic penetration depth $\lambda_{ab}(0)$
\cite{Zhao95,Zhao97,Zhao98,Hofer00,Khasanov03b,Khasanov03,Khasanov04,
Keller03,Keller05,Khasanov04a,Tallon05} poses a challenge to the
understanding of high temperature superconductivity. To date, most
isotope effect studies on $T_c$ and $\lambda_{ab}(0)$ in HTS were
performed by substituting oxygen $^{16}$O with $^{18}$O. It was
observed that the oxygen isotope ($^{16}$O/$^{18}$O) effect (OIE) on
both $T_c$ and $\lambda_{ab}(0)$ have a tendency to increase with
decreasing
doping.\cite{Zhao95,Zhao97,Zhao98,Hofer00,Khasanov03b,Khasanov03,Khasanov04,
Keller03,Khasanov04a,Tallon05,Keller05} Later on it was shown that
for different families of HTS cuprates there is an universal
correlation between the isotope shifts of these two
quantities.\cite{Khasanov03b,Tallon05,Keller03,Khasanov04a,Keller05}
Namely, in the underdoped region $\Delta T_c/T_c$ and
$\Delta\lambda_{ab}(0)/\lambda_{ab}(0)$ scale linearly with respect
to each other with $|\Delta
T_c/T_c|\simeq|\Delta\lambda_{ab}(0)/\lambda_{ab}(0)|$. However,
close to optimal doping the situation is not so clear. Khasanov and
coauthors [\onlinecite{Khasanov04,Khasanov04a}] observed that in
optimally doped YBa$_2$Cu$_3$O$_{7-\delta}$ the small OIE on $T_c$
is associated with a rather big isotope shift of $\lambda_{ab}$ that
is even compatible with the OIE on $\lambda_{ab}$ in underdoped
cuprates. In contrast, Tallon {\it et al.} [\onlinecite{Tallon05}]
showed that in slightly overdoped
La$_{2-x}$Sr$_x$Cu$_{1-y}$Zn$_y$O$_4$ the OIE on $\lambda_{ab}(0)$
is zero while the OIE on $T_c$ remains still substantial.

In this paper we concentrate on studies of the OIE on $T_c$ and
$\lambda_{ab}(0)$ in optimally doped La$_{1.85}$Sr$_{0.15}$CuO$_4$
and YBa$_2$CuO$_{7-\delta}$, as well as in slightly underdoped
YBa$_2$Cu$_4$O$_8$ and Y$_{0.8}$Pr$_{0.2}$Ba$_2$CuO$_{7-\delta}$.
All the samples show a rather small OIE on $T_c$ associated with a
relatively large OIE on $\lambda_{ab}(0)$. The oxygen isotope
exponents on $T_c$ [$\alpha_{\rm O}=-{\rm d}\ln T_c /{\rm d}\ln
M_{\rm O}$, $M_{\rm O}$ is the mass of the oxygen isotope] and the
in-plane magnetic penetration depth $\lambda_{ab}(0)$ [$\beta_{\rm
O}=-{\rm d}\ln \lambda_{ab}(0) /{\rm d}\ln M_{\rm O}$] were found to
be $\alpha_{\rm O}=0.024(8)$ and $\beta_{\rm O}=-0.21(4)$ for
YBa$_2$CuO$_{7-\delta}$, $\alpha_{\rm O}=0.10(1)$ and $\beta_{\rm
O}=-0.19(6)$ for Y$_{0.8}$Pr$_{0.2}$Ba$_2$CuO$_{7-\delta}$,
$\alpha_{\rm O}=0.048(8)$ and $\beta_{\rm O}=-0.18(6)$ for
YBa$_2$Cu$_4$Cu$_4$O$_8$, and $\alpha_{\rm O}=0.08(1)$ and
$\beta_{\rm O}=-0.18(5)$ for La$_{1.85}$Sr$_{0.15}$CuO$_4$. The fact
that a substantial OIE on $\lambda_{ab}(0)$ is observed even in
cuprates having a relatively small OIE on $T_c$ suggests that
lattice effects have to be considered in any realistic model of
high-temperature superconductivity.

\section{Experimental details} \label{sec:experimental_details}

Powder samples of Y$_{1-x}$Pr$_x$Ba$_2$Cu$_3$O$_{7-\delta}$,
YBa$_2$Cu$_4$O$_{8}$ and La$_{1.85}$Sr$_{0.15}$CuO$_4$ were
synthesized by solid state reactions.\cite{Conder01,Karpinski89}
Oxygen isotope exchange was performed during heating the samples in
$^{18}$O$_2$ gas. In order to ensure that the $^{16}$O and $^{18}$O
substituted samples are the subject of the same thermal history, the
annealing of the two samples is performed simultaneously in
$^{16}$O$_2$ and $^{18}$O$_2$ (95\% enriched ) gas, respectively.
The $^{18}$O content of the samples, as determined from a change of
the sample weight after the isotope exchange, were found to be 90\%
for YBa$_2$Cu$_3$O$_{7-\delta}$, 82\% for
Y$_{0.8}$Pr$_{0.2}$Ba$_2$Cu$_3$O$_{7-\delta}$ and
YBa$_2$Cu$_4$O$_8$, and 85\% for La$_{1.85}$Sr$_{0.15}$CuO$_4$. The
total oxygen content for the YBa$_2$Cu$_3$O$_{7-\delta}$ and
La$_{1.85}$Sr$_{0.15}$CuO$_4$ samples  was determined by means of
high-accurate volumetric analysis.\cite{Conder01} The oxygen contens
are: 6.951(2)/6.953(2) and 3.9981(3)/3.9976(3) for the
$^{16}$O/$^{18}$O substituted YBa$_2$Cu$_3$O$_{7-\delta}$ and
La$_{1.85}$Sr$_{0.15}$CuO$_4$ samples, respectively.

In order to determine the OIE on $T_c$, field--cooled magnetization
($M_{FC}$) measurements were performed with a SQUID magnetometer in
a field of 1~mT at temperatures between $1.75$~K and $100$~K.
For the investigation of the OIE on $\lambda_{ab}(0)$,
transverse-field $\mu$SR experiments were performed at the Paul
Scherrer Institute (PSI), Switzerland, using the $\pi$M3 $\mu$SR
facility. The samples were field-cooled from far above $T_{c}$ in a
field of 0.2~T. In a powder sample the magnetic penetration depth
$\lambda$ can be extracted from the muon-spin depolarization rate
$\sigma(T) \propto 1/\lambda^{2}(T)$, which probes the second moment
$\langle \Delta B^{2}\rangle^{1/2}$ of the probability distribution
of the local magnetic field function $p(B)$ in the mixed
state.\cite{Zimmermann95} For highly anisotropic layered
superconductors (like cuprate superconductors) $\lambda$ is mainly
determined by the in-plane penetration depth
$\lambda_{ab}$:\cite{Zimmermann95} $\sigma(T) \propto
1/\lambda_{ab}^{2}(T)$.
The depolarization rate $\sigma$ was extracted from the $\mu$SR time
spectra using a Gaussian relaxation function $R(t) =
\exp[-\sigma^{2}t^{2}/2]$. The superconducting contribution
$\sigma_{sc}$ was obtained by subtracting the dipolar contribution
$\sigma_{nm}$ measured above $T_{c}$ as
$\sigma_{sc}^2=\sigma^2-\sigma_{nm}^2$.

\section{Results and Discussion}

\begin{figure}[htb]
%\centering
\includegraphics[width=1.05\linewidth]{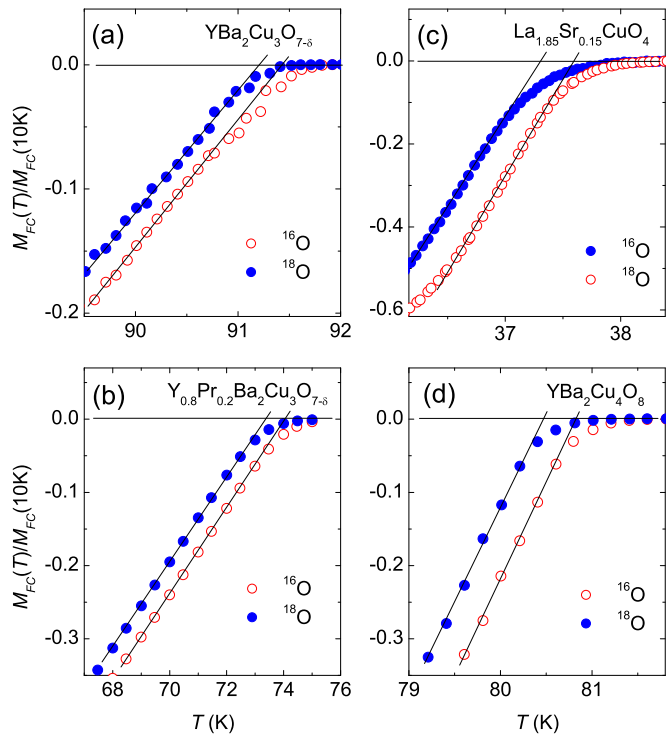}
%\vspace {-1.5cm}
 \caption{(Color online) Section near $T_c$ of the low-field (1mT,
field-cooled) magnetization curves (normalized to the value at 10K)
for $^{16}$O/$^{18}$O substituted YBa$_2$Cu$_3$O$_{7-\delta}$ (a),
Y$_{0.8}$Pr$_{0.2}$Ba$_2$Cu$_3$O$_{7-\delta}$ (b),
La$_{1.85}$Sr$_{0.15}$CuO$_4$ (c), and YBa$_2$Cu$_4$O$_8$ (d).}
 \label{fig:Tc}
\end{figure}

Figure~\ref{fig:Tc} shows the temperature dependences of the
field-cooled magnetization for the $^{16}$O/$^{18}$O substituted
samples investigated in this work. It is seen that the magnetization
curves for the $^{18}$O substituted samples are shifted almost
parallel to lower temperatures, implying that the $T_c$'s of the
$^{18}$O samples are lower than those of the $^{16}$O samples.
The results of the OIE on $T_{c}$ are summarized in
Table~\ref{table:Tc}. The transition temperature $T_c$ was
determined as the temperature where the linearly extrapolated
transition slope intersects the zero line. The OIE exponent
$\alpha_{\rm O}$ is defined by $\alpha_{\rm O}= - d\ln T_c/d\ln
M_{\rm O}$. Taking into account an uncomplete oxygen isotope
exchange (90\% for YBa$_2$Cu$_3$O$_{7-\delta}$, 82\% for
Y$_{0.8}$Pr$_{0.2}$Ba$_2$Cu$_3$O$_{7-\delta}$ and
YBa$_2$Cu$_4$O$_8$, and 85\% for La$_{1.85}$Sr$_{0.15}$CuO$_4$), we
found $\alpha_{\rm O}$ = 0.024(8), 0.10(1), 0.048(8), 0.08(1) for
YBa$_2$Cu$_3$O$_{7-\delta}$,
Y$_{0.8}$Pr$_{0.8}$Ba$_2$Cu$_3$O$_{7-\delta}$, YBa$_2$Cu$_4$O$_8$,
and La$_{1.85}$Sr$_{0.15}$CuO$_4$, respectively. Note that, these
values are in fair agreement with the previously published
results.\cite{Franck91,Franck94,Zech94,Zhao95,Zhao97,Zhao98,Hofer00,
Khasanov03b,Khasanov03,Khasanov04,Keller03,Keller05,Khasanov04a,Tallon05}

\begin{table}
\caption[~]{Summary of the OIE results on $T_c$ for
Y$_{1-x}$Pr$_x$Ba$_2$Cu$_3$O$_{7-\delta}$ ($x=0.0$, 0.2),
YBa$_2$C$_4$O$_8$, and La$_{1.85}$Sr$_{0.15}$CuO$_4$. The values of
$\Delta T_c/T_c$ and $\alpha_{\rm O}$ are corrected for the
uncomplete $^{18}$O exchange (see text for an explanation).
 } %
 \label{table:Tc}
\begin{center}
\begin{tabular}{lllclccc} \hline\hline
   &{$^{16}$O} &{$^{18}$O} &&&
& \\
\hline Sample &$T_c$& $T_c$ &$\Delta T_c/T_c$& $\ \ \alpha_{\rm
O}$\\
 &[K]& [K] &[\%]& &\\
\hline
YBa$_2$Cu$_3$O$_{7-\delta}$ &91.45(5)& 91.20(5) &-0.3(1) &0.024(8) \\
Y$_{0.8}$Pr$_{0.2}$Ba$_2$Cu$_3$O$_{7-\delta}$ &74.0(1)& 73.2(1) &-1.3(3) &0.104(12) \\
YBa$_2$Cu$_4$O$_8$ &80.86(5) &80.46(5) &-0.6(1)&0.048(8)\\
La$_{1.85}$Sr$_{0.15}$CuO$_4$ &37.63.(2)&37.31(2) &-1.0(1)  &0.08(1) \\
\hline \hline
\end{tabular}

\end{center}
\end{table}

\begin{figure}[htb]
%\centering
\includegraphics[width=1.05\linewidth]{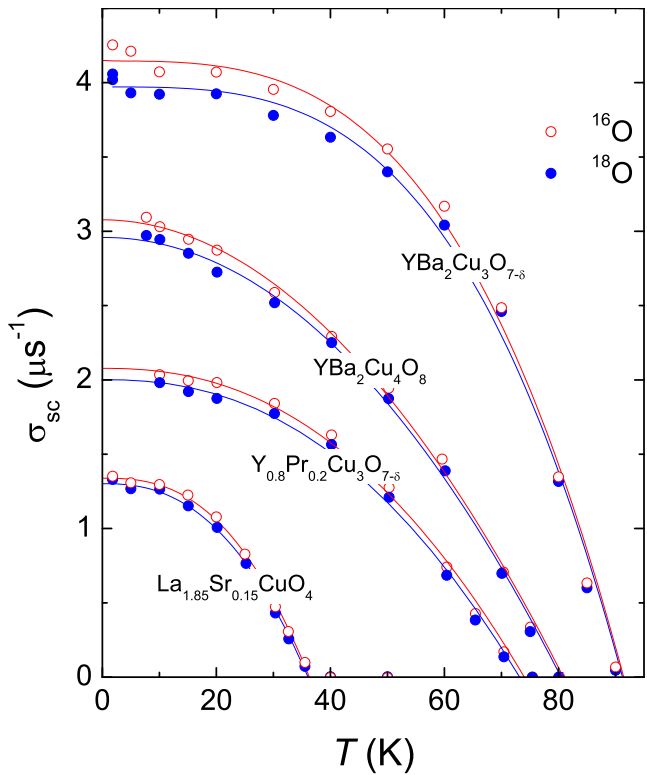}
%\vspace{-0.4cm}
 \caption{(Color online) Temperature dependences of depolarization
rate $\sigma_{sc}$ for $^{16}$O/$^{18}$O substituted (from the top
to the bottom) YBa$_2$Cu$_3$O$_{7-\delta}$, YBa$_2$Cu$_4$O$_8$,
Y$_{0.8}$Pr$_{0.2}$Ba$_2$Cu$_3$O$_{7-\delta}$, and
La$_{1.85}$Sr$_{0.15}$CuO$_4$ samples (200~mT, field-cooled). The
solid lines correspond to fits to the power law
$\sigma_{sc}(T)/\sigma_{sc}(0)=1- (T/T_{c})^n$. The error bars are
smaller than the size of the points.}
 \label{fig:Sigma}
\end{figure}

Figure~\ref{fig:Sigma} shows the temperature dependences of the
superconducting part of the $\mu$SR depolarization rate
$\sigma_{sc}\propto\lambda_{ab}^{-2}$ of the samples studied. It is
seen that the data points for the $^{16}$O substituted samples are
systematically higher than those for the $^{18}$O ones, implying
that an oxygen isotope shift on $\sigma_{sc}$ is present.
As in Ref.~[\onlinecite{Zimmermann95}], the data in
Fig.~\ref{fig:Sigma} were fitted to the power law
$\sigma_{sc}(T)/\sigma_{sc}(0)=1- (T/T_{c})^n$  with
$\sigma_{sc}(0)$ and $n$ as free parameters, and $T_c$ taken from
the magnetization measurements (see Table~\ref{table:Tc}). The
values of $\sigma_{sc}(0)$ obtained from the fits are listed in
Table~\ref{table:sigma} and are in agreement with previous
results.\cite{Seaman90,Zimmermann95,Aepli87} The exponents $n$ were
found to be the same within error for each set of $^{16}$O/$^{18}$O
samples, implying that $\sigma_{sc}$ has nearly the same temperature
dependence for the two isotopes (see Fig.~\ref{fig:Sigma}). The
values of the relative shift of $\lambda_{ab}(0)$ and the oxygen
isotope exponent $\beta_{\rm O}$ obtained from the measured values
of $\sigma_{sc}(0)$ and corrected for the uncomplete $^{18}$O
exchange are summarized in Table~\ref{table:sigma}.

\begin{table}
\caption[~]{Summary of the OIE results on $\lambda_{ab}^{-2}(0)$
for Y$_{1-x}$Pr$_x$Ba$_2$Cu$_3$O$_{7-\delta}$ ($x=0.0$, 0.2),
YBa$_2$Cu$_4$O$_8$, and La$_{1.85}$Sr$_{0.15}$CuO$_4$. The values
of $\Delta \lambda_{ab}(0)/\lambda_{ab}(0)$ and $\beta_{\rm O}$
are corrected for the
uncomplete $^{18}$O exchange (see text for an explanation).  } %
\begin{center}
\begin{tabular}{lllccccc} \hline\hline
   &$^{16}$O &$^{18}$O &&
& \\
\hline Sample &$\sigma(0)$&$\sigma(0)$&$\frac{\Delta
\lambda_{ab}(0)}{\lambda_{ab}(0)}$& $\beta_{\rm
O}$\\
 &[$\mu s^{-1}$]& [$\mu s^{-1}$] &[\%]&  &&\\
\hline
YBa$_2$Cu$_3$O$_{7-\delta}$ &4.15(4)& 3.96(4) &2.6(5)& -0.21(4)  \\
Y$_{0.8}$Pr$_{0.2}$Ba$_2$Cu$_3$O$_{7-\delta}$ &2.08(1)& 2.00(1) &2.4(7)& -0.19(6)  \\
YBa$_2$Cu$_4$O$_8$ &3.07(3)&2.96(3)&2.2(7) &-0.18(6)\\
La$_{1.85}$Sr$_{0.15}$CuO$_4$ &1.34(1)& 1.29(1) &2.2(6)& -0.18(5) \\
\hline \hline
 \label{table:sigma}
\end{tabular}

\end{center}
\end{table}

Note that, the observed OIE's on $T_c$ and $\lambda_{ab}(0)$ are not
caused by a possible difference in the carrier concentrations of the
$^{16}$O and $^{18}$O samples. This is because the oxygen contents
in the $^{16}$O/$^{18}$O substituted YBa$_2$Cu$_3$O$_{7-\delta}$ and
La$_{1.85}$Sr$_{0.5}$CuO$_4$ are the same within error (see
Sec.~\ref{sec:experimental_details}) and YBa$_2$Cu$_4$O$_8$ is a
stoichiometric compound with a fixed oxygen
content.\cite{Karpinski89,Bucher89} Additional arguments are given
in
Refs.~[\onlinecite{Zhao95,Zhao97,Zhao98,Hofer00,Khasanov03b,Khasanov03,Khasanov04,
Keller03,Khasanov04a,Keller05}].

In order to demonstrate that the change of the oxygen content within
the precision of our volumetric analysis cannot account for the
observed OIE on $T_c$ and $\lambda_{ab}(0)$ we used the following
procedure.
Tallon {\it et al.} [\onlinecite{Tallon03}] observed that in a wide
range of doping ($0.05\lesssim p \lesssim 0.19$) the following
empirical relation holds: $T_c\lambda_{ab}^{-2}(0)\propto p-0.05$
($p$ is the number of holes per planar Cu). This implies that
\begin{equation}
\frac{\Delta p}{p-0.05}=\frac{\Delta T_c}{T_c}
-2\frac{\Delta\lambda_{ab}(0)}{\lambda_{ab}(0)}.
 \label{eq:Lambda-Tc}
\end{equation}
Taking into account that oxygen is divalent and that the unit cell
of YBa$_2$Cu$_3$O$_{7-\delta}$ contains two plane and one chain Cu
atoms, the change in the hole concentration $\Delta p$ caused by the
change of the oxygen content can be estimated as $\Delta p=-2\Delta
y$ for La$_{2-x}$Sr$_x$CuO$_{4-y}$ and $\Delta
p=-2/3\cdot\Delta\delta$ for YBa$_2$Cu$_3$O$_{7-\delta}$. For the
above mentioned errors in the determination of the oxygen content
($\pm$0.002 for YBa$_2$Cu$_3$O$_{7-\delta}$ and $\pm$0.0003 for
La$_{1.85}$Sr$_{0.15}$CuO$_4$, see
Sec.~\ref{sec:experimental_details}) and with $p\simeq 0.15$ for the
optimally doped samples, one gets $|\Delta T_c/T_c-2\cdot
\Delta\lambda_{ab}(0)/\lambda_{ab}(0)| < 1.3$\% and $<$0.6\% for
YBa$_2$Cu$_3$O$_{7-\delta}$ and La$_{1.85}$Sr$_{0.15}$CuO$_4$,
respectively. These values are more than five times smaller than one
would obtain by substituting $\Delta T_c/T_c$ and
$\Delta\lambda_{ab}(0)/\lambda_{ab}(0)$ in Eq.~(\ref{eq:Lambda-Tc})
by the values listed in Tables~\ref{table:Tc} and \ref{table:sigma}.

\begin{figure}[htb]
%\centering
\includegraphics[width=1.05\linewidth]{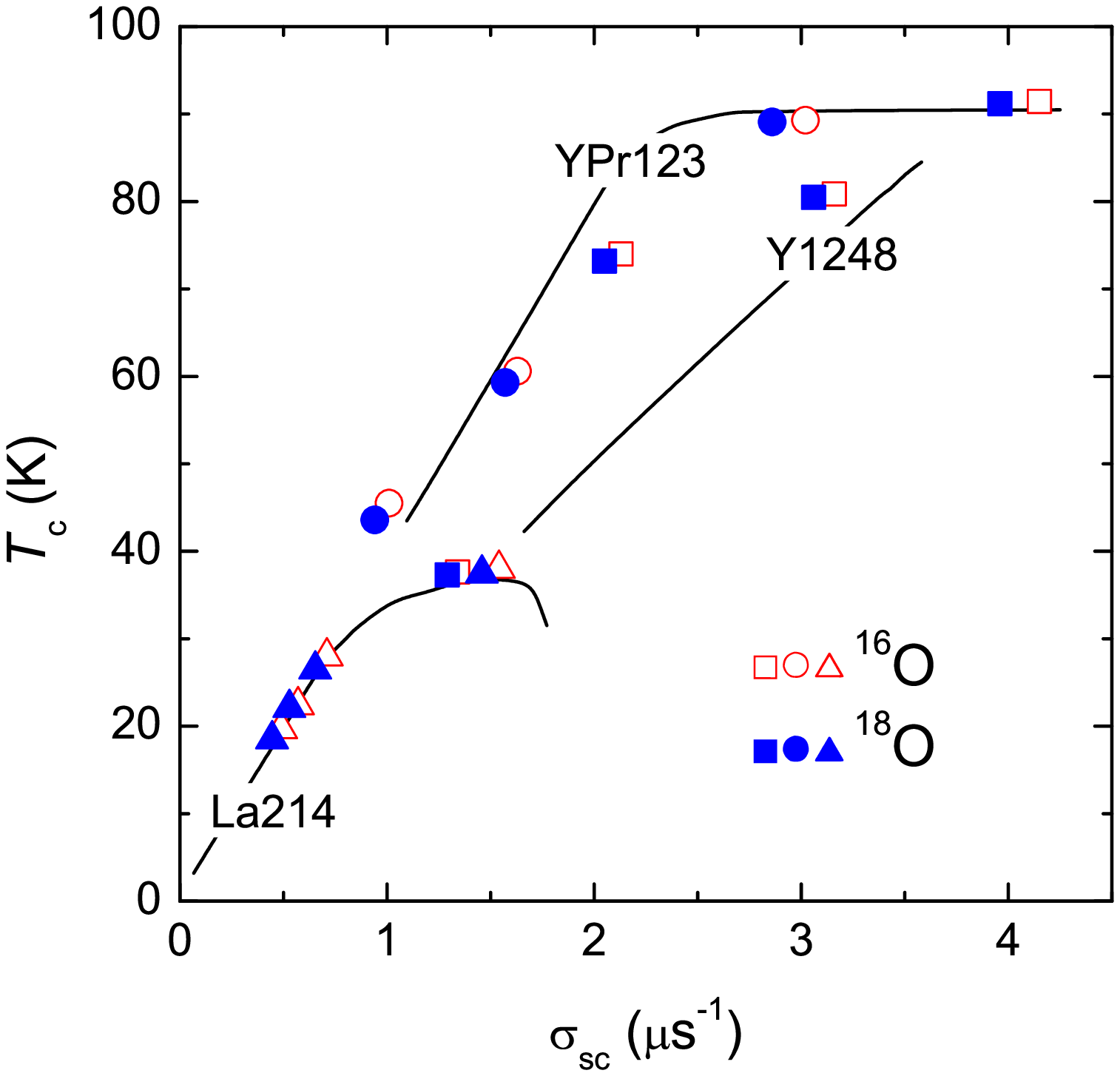}
%\vspace{-0.4cm}
 \caption{(Color online) Plot of $T_c$ versus $\sigma_{sc}(0)$ for
$^{16}$O (open symbols) and $^{18}$O (closed symbols) substituted
Y$_{1-x}$Pr$_{x}$Ba$_2$Cu$_3$O$_{7-\delta}$, YBa$_2$Cu$_4$O$_8$
and La$_{2-x}$Sr$_{x}$CuO$_{4}$. Squares are the $\mu$SR data
obtained in the present study. Circles are bulk $\mu$SR data for
Y$_{1-x}$Pr$_{x}$Ba$_2$Cu$_3$O$_{7-\delta}$
\cite{Khasanov03,Khasanov03b} and LE$\mu$SR data for optimally
doped YBa$_2$Cu$_3$O$_{7-\delta}$.\cite{Khasanov04} Triangles are
torque magnetization and Meissner fraction data for
La$_{2-x}$Sr$_x$CuO$_4$.\cite{Hofer00,Zhao97} The solid lines
correspond to the ''universal`` $T_c$ vs. $\sigma_{sc}(0)$
relations for YPr123, La214 and Y124 families of HTS.
\cite{Uemura89,Uemura91,Tallon95,Bernhard95}}
 \label{fig:uemura}
\end{figure}

\begin{figure}[htb]
%\centering
\includegraphics[width=1.05\linewidth]{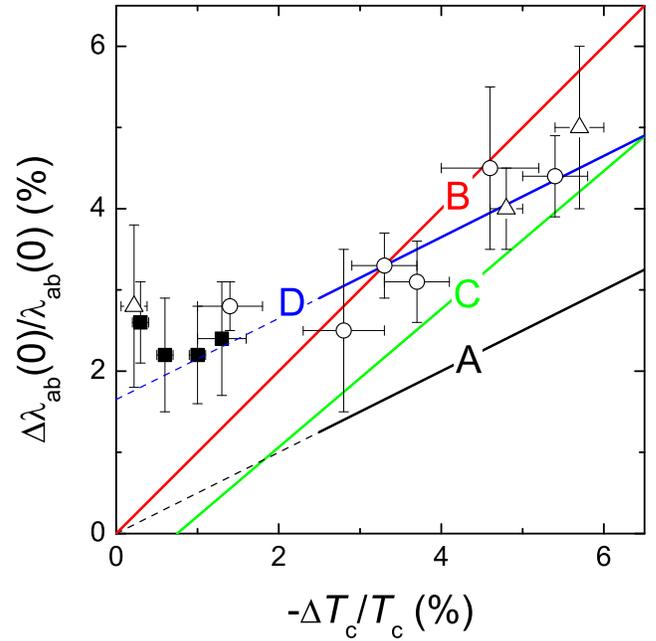}
\vspace{-0.4cm}
 \caption{(Color online) Plot of the OIE shift $\Delta \lambda_{ab}(0)/\lambda_{ab}$(0)  versus
the OIE shift $ - \Delta T_c/T_c$  for
Y$_{1-x}$Pr$_{x}$Ba$_2$Cu$_3$O$_{7-\delta}$, YBa$_2$Cu$_4$O$_8$ and
La$_{2-x}$Sr$_{x}$CuO$_{4}$. Squares are the $\mu$SR data obtained
in the present study. Circles are bulk $\mu$SR data for
Y$_{1-x}$Pr$_{x}$Ba$_2$Cu$_3$O$_{7-\delta}$\cite{Khasanov03,Khasanov03b}
and LE$\mu$SR data for optimally doped
YBa$_2$Cu$_3$O$_{7-\delta}$.\cite{Khasanov04}  Triangles are torque
magnetization and Meissner fraction data for
La$_{2-x}$Sr$_x$CuO$_4$.\cite{Hofer00,Zhao97} The lines correspond
to the ''differential Uemura`` relation with $
\Delta\lambda_{ab}(0)/\lambda_{ab}(0)=0.5|\Delta T_c/T_c|$ (''A``),
$\Delta\lambda_{ab}(0)/\lambda_{ab}(0)=|\Delta T_c/T_c|$ (''B``),
the ''pseudogap`` line from Ref.~[\onlinecite{Tallon05}] (''C``),
and the 2D-QSI relation given by Eq.~(\ref{eq:2D-QSI}) (''D``). The
dashed lines indicates that the ''differential Uemura`` (line ''A``)
and 2D-QSI (line ''D``) relations are strictly valid only in the
underdoped regime (see text for details). }
 \label{fig:alpha-beta}
\end{figure}

In Fig.~\ref{fig:uemura} the transition temperature  $T_c$ is
plotted versus the $\mu$SR depolarization rate
$\sigma_{sc}(0)\propto \lambda_{ab}^{-2}(0)$ for the samples
studied. Recent OIE results for
Y$_x$Pr$_{1-x}$Ba$_2$Cu$_3$O$_{7-\delta}$
\cite{Khasanov03,Khasanov03b,Khasanov04} and
La$_{2-x}$Sr$_x$CuO$_4$ \cite{Zhao97,Hofer00} are also included in
the graph. Since the absolute values of $\lambda_{ab}(0)$ for the
La$_{2-x}$Sr$_x$CuO$_4$ samples studied in
Refs.~[\onlinecite{Zhao97}] and [\onlinecite{Hofer00}] are not
known, the values of $\sigma_{sc}(0)$ for the $^{16}$O substituted
samples were estimated from the comparison with previous
data.\cite{Uemura89,Uemura91}
According to Refs.~[\onlinecite{Uemura89}] and
[\onlinecite{Uemura91}] in the underdoped regime $T_c$ is
proportional to $\sigma_{sc}(0)\propto\lambda_{ab}^{-2}(0)$ with a
universal slope for most HTS families and saturates close to optimal
doping to a value characteristic for each HTS family (Uemura
relation). Furthermore, recent experiments on an ultrathin
La$_{2-x}$Sr$_x$CuO$_4$ film clearly show that the Uemura relation
$T_c\propto n_s/m^\ast$ holds when the superfluid density is
modulated by an electric field.\cite{Rufenacht06} Superconductors
that belong to the Y124 family (including YBa$_2$Cu$_4$O$_8$)
contain CuO chains free of disorder and thus exhibit enhanced values
of $\lambda^{-2}_{ab}(0)$ compared to the "Uemura
line".\cite{Tallon95,Bernhard95} The ''universal`` $T_c$ vs.
$\sigma_{sc}(0)$ curves for the HTS families
Y$_x$Pr$_{1-x}$Ba$_2$Cu$_3$O$_{7-\delta}$ (YPr123),
La$_{2-x}$Sr$_x$CuO$_4$ (La214), and YBa$_2$Cu$_4$O$_8$ (Y124) are
shown in Fig.~\ref{fig:uemura}. Figure~\ref{fig:uemura} suggests
that relation between isotope shifts of $T_c$ and $\lambda_{ab}(0)$
can be explained qualitatively by the empirical ''Uemura line``.
Indeed $T_c$ and $\sigma_{sc}(0)$ for the $^{18}$O substituted
samples are always smaller than those for the $^{16}$O samples. It
is also seen that for samples close to optimal doping a small OIE on
$T_c$ is associated with a rather large OIE on $\lambda_{ab}(0)$.
In order to investigate these results in more detail, we plot in
Fig.~\ref{fig:alpha-beta}  the OIE shift of $\lambda_{ab}(0)$
[$\Delta \lambda_{ab}(0)/\lambda_{ab}(0)$] versus the OIE shift of
$T_c$ ($\Delta T_c/T_c$). It is remarkable that different
experimental techniques (SQUID magnetization,\cite{Zhao97} magnetic
torque,\cite{Hofer00} bulk $\mu$SR,\cite{Khasanov03,Khasanov03b}
low-energy $\mu$SR \cite{Khasanov04}) and different types of samples
(single crystals,\cite{Hofer00}
powders,\cite{Zhao97,Khasanov03,Khasanov03b} thin films
\cite{Khasanov04}) yield consisting results within experimental
error. However, one can easily see that the ''Uemura relation`` can
explain the observed correlation between
$\Delta\lambda_{ab}(0)/\lambda_{ab}(0)$ and $\Delta T_c/T_c$ only
qualitatively but not quantitatively. Following Uemura {\it et al.}
\cite{Uemura89,Uemura91} for different families of underdoped
cuprates, $T_c$ scales linearly with the $\mu$SR depolarization rate
$\sigma_{sc}(0)\propto\lambda_{ab}^{-2}(0)$, yielding $
\Delta\lambda_{ab}(0)/\lambda_{ab}(0)\simeq 0.5 |\Delta T_c/T_c|$
(line ''A`` in Fig.~\ref{fig:alpha-beta}). It is seen, however, that
all the experimental points are systematically higher. At low doping
level $\Delta\lambda_{ab}(0)/\lambda_{ab}(0)\simeq | \Delta
T_c/T_c|$ (line ''B`` in Fig.~\ref{fig:alpha-beta}), whereas close
to the optimal doping $\Delta\lambda_{ab}(0)/\lambda_{ab}(0)$ is
almost constant and considerably larger than $\Delta T_c/T_c$
($\Delta\lambda_{ab}(0)/\lambda_{ab}(0)\approx 10 |\Delta
T_c/T_c|$).\cite{Khasanov04a,Keller03,Keller05}

According to the London theory $\lambda^{-2}_{ab}$ is proportional
to the so called ''superfluid density`` $\lambda^{-2}_{ab}\propto
n_s/m^\ast_{ab}$ ($n_s$ is the density of supercarriers and
$m^\ast_{ab}$ is the in-plane charge carrier mass).
Concerning the relation between
$\Delta\lambda_{ab}(0)/\lambda_{ab}(0)$ and $\Delta T_c/T_c$, one
should distinguish two cases: (i) change of the carrier density $n$
by doping (note that $n$ is not necessarily equal to $n_s$ as
discussed in Ref.~[\onlinecite{Rufenacht06}]), (ii) change of the
oxygen mass by isotope substitution.
In the recent electrostatic modulation experiments,
\cite{Rufenacht06} it was shown that the change of the carrier
density within the {\it same} sample leads to $T_c\propto n_s$, and
$\Delta\lambda_{ab}(0)/\lambda_{ab}(0)=0.5|\Delta T_c/T_c|$, in
accordance with the Uemura relation in the underdoped regime.
\cite{Uemura89,Uemura91} Note that in
Ref.~[\onlinecite{Rufenacht06}] carriers were implanted/removed to
the sample by changing the electric field, so the crystal lattice is
not affected. This implies that in this case change of both $T_c$
and $\lambda_{ab}(0)$ are due to a change of the carrier
concentration $n_s$, while the in-plane charge carrier mass
$m^\ast_{ab}$ stays constant.\cite{Niedermayer93} The isotope
substitution, in contrast, modifies the lattice, but leaves the
doping (oxygen content) unchanged (see discussion above). In
addition, it is found that
$\Delta\lambda_{ab}(0)/\lambda_{ab}(0)=|\Delta T_c/T_c|$, so there
is a factor of two difference in the slope of lines ''A`` and ''B``
in Fig.~\ref{fig:alpha-beta}. This factor two can be explained by a
simple model, assuming that $T_c\propto n_s$ and
$\lambda_{ab}^{-2}(0)\propto n_s/m^\ast_{ab}$ (London model). This
implies that in case (i) described above,
$\Delta\lambda_{ab}(0)/\lambda_{ab}(0)=0.5|\Delta
T_c/T_c|=0.5|\Delta n_s/n_s|$ (Uemura relation). However, in case
(ii) the isotope substitution would lead to a change in $n_s$ as
well as in $m^\ast_{ab}$ ($^{16}m^\ast_{ab}<$$^{18}m^\ast_{ab}$) in
order to account for the factor two.

Now we discuss the observed $\Delta \lambda_{ab}(0)/\lambda_{ab}(0)$
vs. $\Delta T_c/T_c$ dependence presented in
Fig.~\ref{fig:alpha-beta} in more detail.
Tallon {\it et al.} \cite{Tallon05} showed that the relation between
the oxygen isotope shifts of $\lambda_{ab}(0)$ and $T_c$ may be
understand in terms of a normal state pseudogap which competes with
superconductivity. Both $\Delta\lambda_{ab}(0)/\lambda_{ab}(0)$ and
$|\Delta T_c/T_c|$ were found to increase with increasing pseudogap
energy $E_g$. At the critical doping (when the pseudogap is closed)
$\Delta\lambda_{ab}(0)/\lambda_{ab}(0)$ is equal to zero, while
$|\Delta T_c/T_c|\simeq0.8$\% (line ''C`` in
Fig.~\ref{fig:alpha-beta}). This is, however, inconsistent with the
experimental data presented in Fig.~\ref{fig:alpha-beta}, especially
close to optimal doping (see Fig.~\ref{fig:alpha-beta}).
%On the other hand,
Schneider and Keller
\cite{Schneider01,Schneider03,KellerSchneider04,SchneiderKeller04}
showed that the relation between the isotope shifts of
$\lambda_{ab}(0)$ and $T_c$ arises naturally from the doping driven
3D-2D crossover and 2D quantum superconductor to insulator (2D-QSI)
transition in the highly underdoped limit. Close to the 2D-QSI
transition the following relation
holds:\cite{KellerSchneider04,SchneiderKeller04}
\begin{equation}
\Delta\lambda_{ab}(0)/\lambda_{ab}(0)=(\Delta d_s/d_s-\Delta
T_c/T_c)/2,
 \label{eq:2D-QSI}
\end{equation}
where $\Delta d_s/d_s$ is the oxygen isotope shift of the thickness
of the superconducting sheets $d_s$ of the sample. The best fit of
the Eq.~(\ref{eq:2D-QSI}) to the experimental data gives $\Delta
d_s/d_s=3.3(4)$\% (line ''D`` in Fig.~\ref{fig:alpha-beta}). Note
that Eq.~(\ref{eq:2D-QSI}) is strictly valid only in the underdoped
region. The fit, however, describes the behavior at all doping
levels reasonably well.  Since the lattice parameters are not
modified by oxygen substitution \cite{Conder94,Raffa98} the
observation of an isotope effect on $d_s$ implies local lattice
distortions involving oxygen that are coupled to the superfluid. It
was shown that in anisotropic superconductors falling into the
2D-XY-QSI universality class at zero temperature, the isotope
effects on the transition temperature, specific heat and the
magnetic field penetration depth are related by a universal
relation. This implies a dominant role of fluctuations so that pair
formation and pair condensation do not occur simultaneously. From
these Schneider and Keller conclude that the observed isotope
effects do not provide direct information on the underlying pairing
mechanism and must be attributed to the shift of the phase diagram
upon isotope substitution caused by electron-lattice
interaction.\cite{Schneider01,Schneider03,KellerSchneider04,SchneiderKeller04}
Bussmann-Holder {\it et
al.}\cite{Bussmann-Holder05,Bussmann-Holder05a,Bussmann-Holder05b}
investigated the origin of the isotope effects on the
superconducting transition temperature and the magnetic penetration
depth within a polaronic model. For this purpose the well-known {\it
t-J} Hamiltonian was extended to incorporate the hole induced charge
channel and the important effects from the lattice. This results in
a two-component Hamiltonian, where interactions between the charge
channel (local hole plus induced lattice distortion) and the spin
channel (antiferromagnetic fluctuations modified by lattice
distortions) are explicitly
included.\cite{Bussmann-Holder05,Bussmann-Holder05a,Bussmann-Holder05b}
This polaronic model predicts for the OIE on $T_c$ and
$\lambda_{ab}(0)$ the relation
$\Delta\lambda_{ab}(0)/\lambda_{ab}(0)=|\Delta T_c/T_c|$ (line D in
the Fig.~\ref{fig:alpha-beta}), in agreement with experiments in the
underdoped regime.

\section{conclusion}

In conclusion, the oxygen isotope $^{16}$O/$^{18}$O effects on the
in-plane magnetic penetration depth $\lambda_{ab}(0)$ and transition
temperature $T_c$ were studied in optimally doped
YBa$_2$Cu$_3$O$_{7-\delta}$ and La$_{1.85}$Sr$_{0.15}$CuO$_4$, and
in the slightly underdoped YBa$_2$Cu$_4$O$_8$ and
Y$_{0.8}$Pr$_{0.2}$Ba$_2$Cu$_3$O$_{7-\delta}$ by means of muon-spin
rotation and magnetization techniques. A small OIE on the transition
temperature $T_c$ was observed that is associated with a substantial
OIE on the in-plane penetration depth $\lambda_{ab}(0)$, as shown in
Fig.~\ref{fig:alpha-beta} and Tables ~\ref{table:Tc} and
\ref{table:sigma}. This finding suggests that lattice effects are
{\it directly or indirectly} involved in determining the
superconducting state. It is worth to note that in colossal
magnetoresistance (CMR) manganites similar peculiar OIE on various
quantities (e.g. ferromagnetic transition temperature,
charge-ordering temperature) were observed,\cite{Zhao99} indicating
that in both classes of perovskites, HTS and CMR manganites, lattice
vibrations play an essential role.

This work was partly performed at the Swiss Muon Source (S$\mu$S),
Paul Scherrer Institute (PSI, Switzerland). The authors are grateful
to D.~Di~Castro, D.G.~Eshchenko, A.~Amato and D.~Herlach for
assistance during the $\mu$SR measurements. This work was supported
by the Swiss National Science Foundation, in part by the NCCR
program MaNEP, and by the K.~Alex~M\"uller Foundation.

%\newpage

%

\end{document}